\documentclass[aps,prl,twocolumn,preprintnumbers,superscriptaddress,10pt]{revtex4-1}

\pdfoutput=1

\usepackage[a4paper,left=2.73cm,right=2.7cm,top=3cm,bottom=3.5cm]{geometry}

\usepackage[T1]{fontenc} 
\usepackage{graphicx}
\usepackage{epsfig}
\usepackage{amssymb}
\usepackage{amsfonts}
\usepackage{amsmath,euscript,array,mathrsfs}
\usepackage{bbold}
\usepackage{epsf}
\usepackage{slashed}
\usepackage{comment}
\usepackage{colortbl}

\usepackage[caption=false]{subfig}
\usepackage[colorlinks=true,linktocpage=true,linkcolor=blue,citecolor=blue]{hyperref}

\usepackage{tikz}
\usetikzlibrary{decorations.markings,decorations.pathmorphing}

\newcommand{\be}{\begin{equation}}
\newcommand{\ee}{\end{equation}}
\newcommand{\beqs}{\begin{eqnarray}}
\newcommand{\eeqs}{\end{eqnarray}}

\newcommand{\eqn}[1]{(\ref{#1})}
\newcommand{\fig}[1]{Fig.~(\ref{#1})}

\newcommand{\sac}{\, , \qquad}

\begin{document}
\onecolumngrid

%\preprint{ICCUB-21-004}

\title{Quantum Matter near a Cosmological Singularity}

\author{Jorge Casalderrey-Solana}
\affiliation{Departament de F\'\i sica Qu\`antica i Astrof\'\i sica \&  Institut de Ci\`encies del Cosmos (ICC), Universitat de Barcelona, Mart\'{\i}  i Franqu\`es 1, 08028 Barcelona, Spain}
\author{David Mateos}
\affiliation{Departament de F\'\i sica Qu\`antica i Astrof\'\i sica \&  Institut de Ci\`encies del Cosmos (ICC), Universitat de Barcelona, Mart\'{\i}  i Franqu\`es 1, 08028 Barcelona, Spain}
\affiliation{Instituci\'o Catalana de Recerca i Estudis Avan\c cats (ICREA), %Passeig 
Llu\'\i s Companys 23, Barcelona, Spain}
\author{Alexandre Serantes}
\affiliation{Departament de F\'\i sica Qu\`antica i Astrof\'\i sica \&  Institut de Ci\`encies del Cosmos (ICC), Universitat de Barcelona, Mart\'{\i}  i Franqu\`es 1, 08028 Barcelona, Spain}
\affiliation{Department of Physics and Astronomy, Ghent University, 9000 Ghent, Belgium}

%\pacs{11.25.Tq, 12.38.Mh}

\begin{abstract}
General Relativity predicts that the spacetime near a cosmological singularity undergoes an infinite number of chaotic oscillations between different Kasner epochs with rapid transitions between them. This so-called BKL behaviour persists in the presence of several types of classical matter. Little is known in the presence of quantum effects. A major obstacle is the fact that the fast metric oscillations inevitably drive the matter far from equilibrium. We use holography to determine the evolution of the quantum stress tensor of a non-conformal, strongly-coupled, four-dimensional gauge theory in a Kasner spacetime. The stress tensor near the singularity is solely controlled by the ultraviolet fixed point of the gauge theory, and it diverges in a universal way common to all theories with a gravity dual. We then compute the backreaction of the  stress tensor on the Kasner metric to leading order in the gravitational coupling. The modification of the Kasner exponents that we find suggests that the BKL behaviour may be avoided in the presence of quantum matter.
\end{abstract}


\maketitle

%%%%%%%%%%%%%%%%%%%%%%%%%%%%%%%%%%%%%%%%%%%%%%%%
\noindent
{\bf 1. Introduction.} 
%\label{intro}
%%%%%%%%%%%%%%%%%%%%%%%%%%%%%%%%%%%%%%%%%%%%%%%%
Singularities such as those in the interior of a black hole, or in a cosmological ``Big Bang'' or ``Big Crunch'', are one of the most profound predictions of Einstein's theory of General Relativity (GR) and, at the same time, a signal of its own breakdown. In the late 1950's Landau stated that they are one of the three most important problems in theoretical physics \cite{Khalatnikov:2008zt}. Enormous progress has been made since then in the presence of purely classical dynamics. GR predicts that the spacetime undergoes the so-called BKL behaviour \cite{Belinsky:1970ew,Belinski:2017fas}, namely chaotic dynamics whereby an infinite series of Kasner epochs take place before the singularity is reached, with rapid transitions between them. This result persists in the presence of several types of classical matter \cite{Belinski:1973zz,YM_fields,Damour:2002tc}.
%\footnote{Exceptions include a massless scalar field or a perfect fluid with equation of state $E=P$, but these are considered non-generic, since addition of other forms of  matter restores the BKL behaviour.} 
The  BKL result is considered a major achievement in GR because it provides a complete physical picture of the dynamics of spacetime near a cosmological singularity.

A fundamental open question is what happens in the presence of quantum effects. While understanding the ultimate fate of the singularity may involve quantum gravity, there is an  interesting regime in which gravity remains classical but quantum field theoretical (QFT) effects, henceforth quantum matter effects,  are important. Despite this simplification, this semiclassical regime is widely applicable. Specifically, it is expected to be relevant provided the energy densities and curvatures are smaller than the so-called species cut-off \cite{Dvali:2007hz}
\be
\label{species}
M_{\rm{sp}} \equiv M_{\rm{P}}/N \,, 
\ee
with $M_{\rm{P}}$ the Planck scale and $N^2$ the number of matter degrees freedom \footnote{Holographic studies  are consistent with \eqn{species} \cite{Chesler:2020exl,Ghosh:2023gvc}. For an extensive discussion in string theory see \cite{Castellano:2022bvr}.}. In this paper we will be interested in a large-$N$ gauge theory, so we will work in the limit $M_{\rm{P}}, N \to \infty$ with $M_{\rm{sp}}$ fixed. In terms of Newton's constant $G$, this limit corresponds to $G\to 0, N\to \infty$ with 
\be
\mathcal{G}\equiv G N^2 = 1/ 8\pi M_{\rm{sp}}^{2} 
\ee
fixed. It is then convenient to work with a rescaled gauge theory stress tensor defined through 
\be
T^\mu_{\, \nu}=\frac{N^2}{2\pi^2} \left( -\mathcal{E}, \mathcal{P}_1, \mathcal{P}_2, \mathcal{P}_3 \right) \,.
\ee

Semiclassical gravity near a cosmological singularity is still extremely challenging because the large gradients inevitably drive the matter far from equilibrium. QFT in this regime is remarkably hard to analyse with conventional methods in flat space, let alone in a dynamical spacetime (see e.g.~\cite{Birrell:1982ix}).
%For this reason, conventional approaches must restrict themselves to free theories \cite{Zeldovich:1970si,Zeldovich:1971mw,Zeldovich:1977vgo} or resort to unjustified approximations such as adiabaticity \cite{Matyjasek:2018tro}, truncations \cite{Alonso-Serrano:2021ufj}, etc. 
For this reason, in this paper we will resort to holography, also known as gauge/gravity duality. Holography maps the QFT problem in four dimensions to a classical gravitational problem in five dimensions, hence allowing for the study of the dynamics arbitrarily far from equilibrium. 

We will use holography to determine the stress tensor of a non-conformal, strongly-coupled, four-dimensional gauge theory in the simplest cosmological singularity, namely in a Kasner metric of the form
\begin{equation}
\label{dh2_K}
ds^2_{\rm{K}} = - dt^2 + \sum_{i=1}^3 a_i(t)^2 \, dx_i^2 \,, \,\,\,\,\,\, a_i(t)= |t|^{p_i}\,.
%|t|^{2 p_x} dx^2 + |t|^{2 p_y} dy^2 + |t|^{2 p_z} dz^2 \,.     
\end{equation}
In this metric $t$ is the proper time to the singularity, which is located at $t=0$ and we imagine approaching from negative $t$, and the Kasner exponents obey
\begin{equation}
\label{exponentssum}
 \sum_{i=1}^3 p_i=1 \sac  \sum_{i=1}^3 p_i^2=1 \,.
\end{equation}
The Hubble or expansion rates in each direction are given by 
\be
\label{exp}
H_i=\frac{\dot a_i}{a_i}=\frac{p_i}{t}\,,
\ee
and the curvature scales as 
\be
\label{curv}
\mathcal{R} \sim t^{-2}\,.
\ee 

We will see that the gauge theory is driven arbitrarily far from equilibrium as the singularity is approached. In particular, the energy density diverges in this limit as 
$\mathcal{E}\propto t^{-2}$. As a consequence, the stress tensor near the singularity is solely controlled by the ultraviolet (UV) fixed point of the gauge theory. Moreover, in this limit the dual bulk geometry simplifies to a universal form, suggesting that the divergence of the stress tensor is universal for any gauge theory with a gravity dual. After determining the stress tensor in a fixed Kasner geometry, we will compute its perturbative backreaction on the Kasner metric to leading order in $\mathcal{G}$. We will find that this modifies the right-hand side of eqs.~\eqn{exponentssum} in a way that suggests that the BKL behaviour may be avoided. 

In this paper we will describe the physical results in the simplest possible way. Further details  will be presented in \cite{prep}.

%%%%%%%%%%%%%%%%%%%%%%%%%%%%%%%%%%%%%%%%%%%%%%%%
\noindent
{\bf 2. Set-up.}
%\label{setup}
%%%%%%%%%%%%%%%%%%%%%%%%%%%%%%%%%%%%%%%%%%%%%%%%
\begin{figure}[t]
\center
\includegraphics[width=0.49\textwidth]{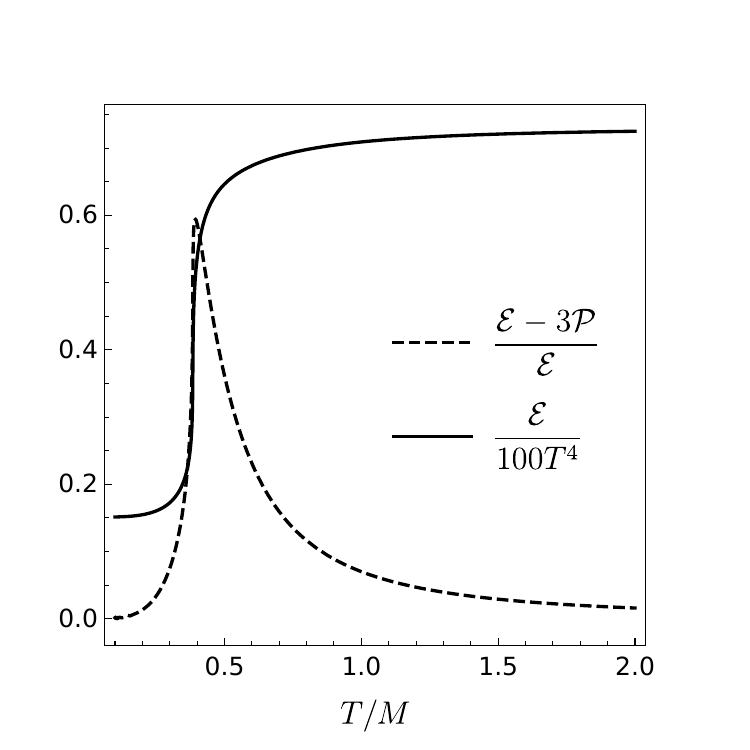}
\caption{\small\small Energy density and interaction measure of the gauge theory.}
\label{eos}
\end{figure}
We consider the model introduced in \cite{Bea:2018whf} in the case in which the gauge theory describes a renormalization group flow between an UV and an infrared (IR) fixed point. The crossover between the two takes place at a characteristic energy scale $M$. This is illustrated by the thermodynamics of the theory, as shown in \fig{eos}. 
The energy density approaches  $T^4$ at temperatures above and below $M$, as expected from conformal invariance near the fixed points. Similarly, the deviation from conformality, as measured by the so-called interaction measure 
\be
\label{inter}
I=(\mathcal{E}-\sum_{i=1}^3 \mathcal{P}_i)/\mathcal{E} \,,
\ee
approaches zero in these limits.  We will measure all dimensionful quantities, including $t$ and $x_i$ in \eqn{dh2_K},  in units of $M$.

\begin{figure}[t!!!]
\center
\includegraphics[width=0.49\textwidth]{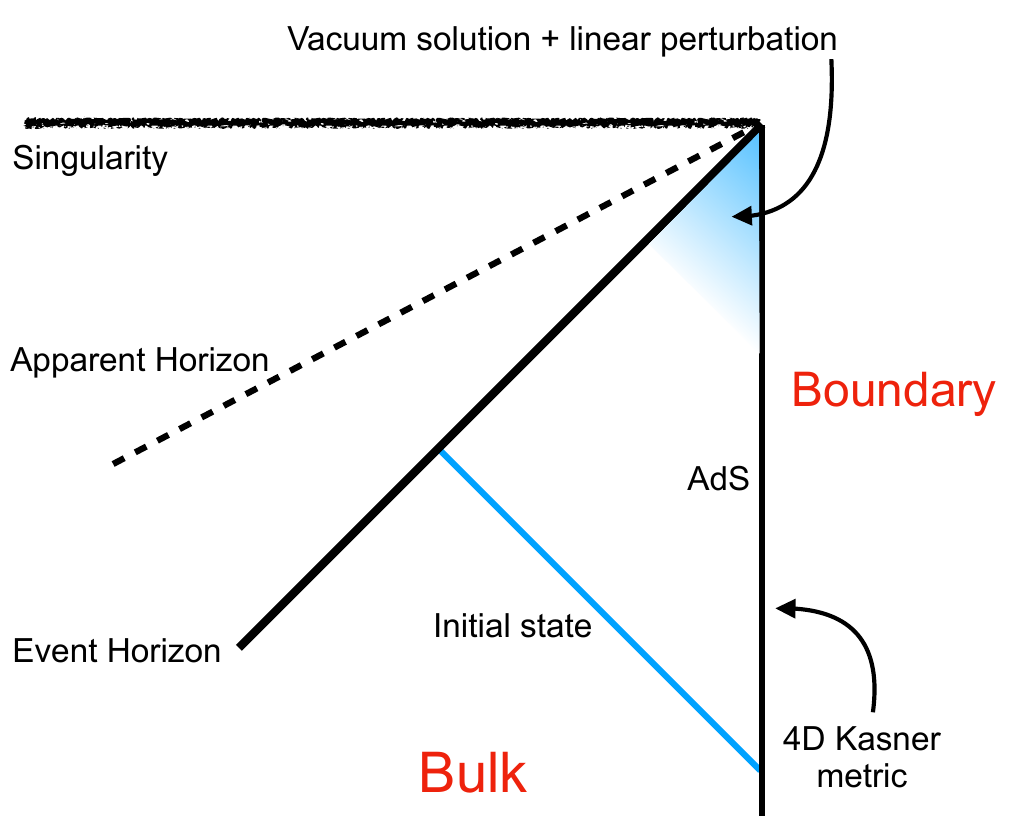}
\caption{\small\small Penrose diagram of the five-dimensional geometry dual to a four-dimensional gauge theory in the Kasner spacetime \eqn{dh2_K}.}
\label{penrosediagram}
\end{figure}

The holographic dual consists of gravity in five dimensions coupled to a scalar field $\phi$. The scalar is subject to a potential such that the solutions are asymptotically locally anti-de Sitter (AdS), as shown in  \fig{penrosediagram}. The energy scale $M$ arises on the gravity side as a boundary condition on $\phi$. Additionally, since we are interested in the behaviour of the gauge theory in a Kasner spacetime, we impose the four-dimensional Kasner metric \eqn{dh2_K} as a boundary condition on the 
 five-dimensional metric.

Since we will use a characteristic formulation of the Einstein plus Klein-Gordon (EKG) equations, we choose to foliate the five-dimensional geometry with null slices, as indicated in \fig{penrosediagram}. We specify the initial state at $M t_{\rm{ini}}=-20$ on the initial slice shown as a blue line in the figure. We then evolve the EKG  equations in time using numerical methods along the lines of \cite{Chesler:2008hg,Chesler:2009cy,Chesler:2010bi,Casalderrey-Solana:2013aba,Chesler:2013lia,Casalderrey-Solana:2013sxa,Chesler:2015wra,Attems:2016tby,Casalderrey-Solana:2016xfq,Attems:2017ezz,Attems:2017zam,Attems:2018gou,Attems:2019yqn,Casalderrey-Solana:2020vls,Bea:2021zsu,Bea:2021ieq,Bea:2021zol,Bea:2022mfb}. From the near-boundary behaviour of the five-dimensional metric we read off the gauge theory stress tensor \footnote{Since \eqn{dh2_K} is Ricci-flat there are no ambiguities associated to curvature-squared finite counterterms.}.

\begin{figure}[t]
\center
\includegraphics[width=0.49\textwidth]{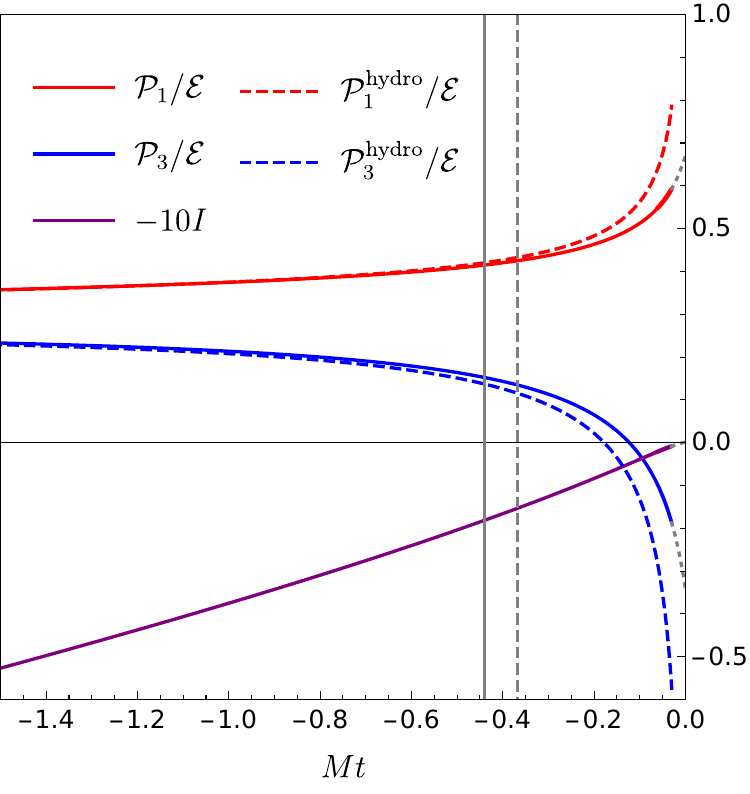}
\caption{\small\small Comparison between the exact pressures (solid curves) and the prediction of first-order, viscous hydrodynamics (dashed curves). The dotted grey curves indicate extrapolation to $t=0$. The solid vertical line indicates the time after which hydrodynamics fails by more than 10\%. The dashed vertical line indicates the time after which the largest expansion rate exceeds the local temperature, i.e.~$H_1>T$. The purple curve shows the deviation from conformality \eqn{inter}.}
\label{hydro}
\end{figure}

%%%%%%%%%%%%%%%%%%%%%%%%%%%%%%%%%%%%%%%%%%%%%%%%
\noindent
{\bf 3. Numerical evolution.}
%\label{evolution}
%%%%%%%%%%%%%%%%%%%%%%%%%%%%%%%%%%%%%%%%%%%%%%%%
For concreteness, in this section we choose a Kasner geometry \eqn{dh2_K} with $p_1=p_2=2/3, p_3=-1/3$ and impose rotational symmetry in the 12-plane, so that the pressures obey $\mathcal{P}_1=\mathcal{P}_2$.  We choose the initial state to be near equilibrium. In particular, its initial temperature $T(t_{\rm{ini}})=0.48 M$ is higher than the instantaneous expansion rates at the initial time, $H_i(t_{\rm{ini}})=0.05 p_i M$. These conditions ensure that the early stages of the evolution are well described by hydrodynamics. This is illustrated in \fig{hydro}, where we compare the exact pressures, normalized to the energy density, with the predictions of first-order, viscous hydrodynamics. The proximity between the two vertical lines in the figure indicates that hydrodynamics begins to fail once the expansion rate becomes larger than the local temperature.   

\begin{figure}[t]
\center
\includegraphics[width=0.49\textwidth]{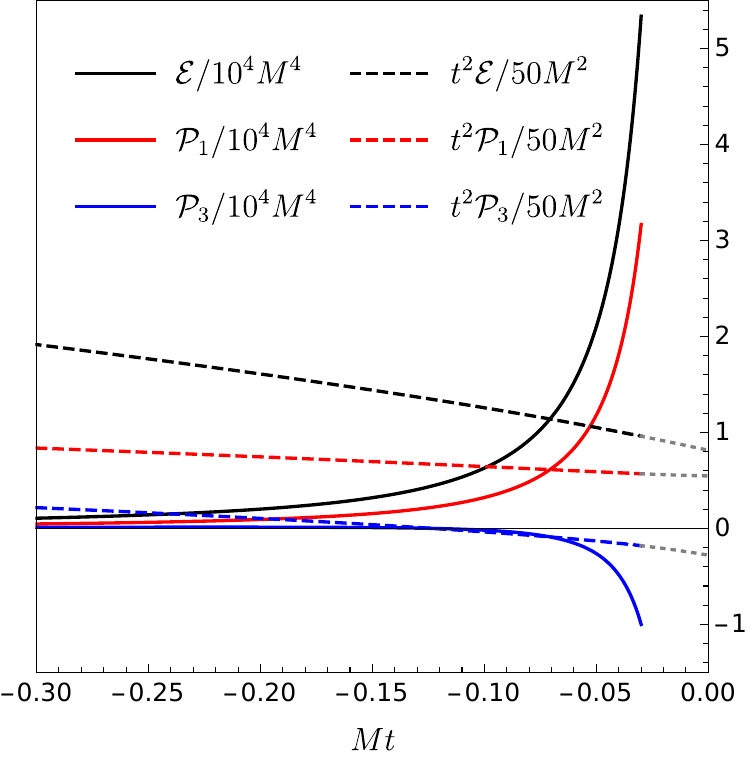}
\caption{\small\small Behaviour of the energy density and pressures near the singularity. The dotted grey curves indicate extrapolation to $t=0$.}
\label{divergence}
\end{figure}

In \fig{hydro} we also show the deviation from conformality \eqn{inter}. The fact that this  approaches zero at $t=0$ indicates that this limit is controlled by the UV fixed point of the theory. In this limit the energy and the pressures diverge as 
\be
\label{divergent}
\mathcal{E} = \frac{\Lambda^2}{t^2} \sac 
\mathcal{P}_i = p_i \, \mathcal{E} \sac 
T \propto \mathcal{E}^{1/4} \,,
\ee
where $\Lambda$ is a constant with dimensions of mass that depends on the initial conditions. This behaviour is illustrated in \fig{divergence}, where we see that $t^2 \mathcal{E}$ and $t^2 \mathcal{P}_i$ approach finite limits at $t=0$. The ratios between the extrapolations of these quantities to $t=0$ are consistent with \eqn{divergent}. By conformal invariance the temperature scales as 
\be
\label{TTT}
T \sim \mathcal{E}^{1/4} \sim \left( \frac{\Lambda}{t} \right)^{1/2} \,.
%\left( \Lambda/t \right)^{1/2}\,.
\ee
This means that higher and higher gradients in the hydrodynamic expansion diverge faster and faster, confirming its complete breakdown.

%%%%%%%%%%%%%%%%%%%%%%%%%%%%%%%%%%%%%%%%%%%%%%%%
\noindent
{\bf 4. Universality.}
%\label{evolution}
%%%%%%%%%%%%%%%%%%%%%%%%%%%%%%%%%%%%%%%%%%%%%%%%
We will now show that the behaviour \eqn{divergent} is solely determined by the UV fixed point of the theory. We begin by noting that, since \eqn{dh2_K} is Ricci-flat, the five-dimensional metric
\be
\label{vacFG}
ds^2_{\rm{5D}} = \frac{ds^2_{\rm{K}} +du^2}{u^2}
\ee
is a solution of the bulk EKG equations provided we set \mbox{$\phi=0$}. In the gauge theory this condition corresponds to setting $M=0$, namely to replacing the entire non-conformal theory by its UV fixed point. Moreover,  the boundary stress tensor associated to the metric \eqn{vacFG} vanishes. This suggests that one should think of \eqn{vacFG} as the ``vacuum solution'' dual to the ``vacuum state'' of the UV theory.

The change of coordinates 
%$r=u^{-1}, \tau = t - u$
\be
%r = \frac{1}{u} \sac \tau = t - u
v=u \sac \tau = t - u 
\ee
puts the metric \eqn{vacFG} in the Eddington-Finkelstein-like (EF) form
\be
\label{vacEF}
%ds^2 = - r^2 d\tau^2 + 2 d\tau dr + \sum_{i=1}^3 r^2 \Big(-\tau-\frac{1}{r} \Big)^{2 p_i} dx_i^2 \,.
ds^2 = \frac{1}{v^2} \Big[ - d\tau^2 - 2 d\tau dv + 
\sum_{i=1}^3 \Big(-\tau-v  \Big)^{2 p_i} dx_i^2 \Big] \,.
\ee
The boundary is at $v=0$, and there $t=\tau$. This metric possesses a curvature singularity on the spacelike hypersurface $v_{\rm{sing}}=|\tau|$, indicated by a horizontal line in \fig{penrosediagram}. The goal of Refs.~\cite{Engelhardt:2013jda,Engelhardt:2014mea,Engelhardt:2015gta} was to use the boundary gauge theory to understand this bulk singularity. Our purpose is the opposite, namely, we want to use the bulk spacetime to understand the behaviour of quantum fields near the $t=0$ singularity  of the boundary metric  \eqn{dh2_K}. The reason this is possible is that the boundary is shielded from the bulk singularity by an event horizon located on the null hypersurface $v_{\rm{EH}}=(1/2)|\tau|$, indicated by a 45-degree line in \fig{penrosediagram}. This is qualitatively different from perhaps-more-familiar cases because here the boundary geometry terminates at a finite proper time, $t=0$. The curve $v=(1/2)|\tau|$ is an outgoing null geodesic in the bulk that reaches the boundary precisely at this time. Colloquially speaking, this is the ``last'' bulk null geodesic that is able to reach the boundary. Bulk points with  $v> (1/2)|\tau|$ are  causally disconnected from the boundary because outgoing null geodesics emanating from them reach the bulk singularity before reaching the boundary. In between the event horizon and the singularity there is also an apparent horizon located at $v_{\rm{AH}}=(3/4)|\tau|$, indicated in \fig{penrosediagram} by a dashed line. The ordering $v_{\rm{sing}}>  v_{\rm{AH}} > v_{\rm{EH}}$ implies that an observer falling from the boundary into the bulk first encounters the EH, then the AH, and finally the singularity. 

As we approach the boundary singularity at $t=\tau=0$,  the range of the EF-coordinate for points causally connected with the boundary shrinks to zero, since $v\leq (1/2)|\tau|$. This suggests that the leading, near-singularity behaviour of arbitrarily excited states, even those with a divergent stress tensor as \eqn{divergent}, can be captured by linearized perturbations around the vacuum solution \eqn{vacEF}. The intuition comes from the fact that the effect of a non-zero stress tensor  scales as $v^4$ near the boundary, and this vanishes for points with $v\leq (1/2)|\tau|$ in the limit $\tau \to 0$. A detailed analysis \cite{prep} shows that linearized perturbations of the vacuum solution behave similarly to the way that quasi-normal modes (QNM) behave in time-independent black hole geometries. In particular, the leading solution near $\tau=0$ is controlled by the analog of the lowest QNM, and its form near the AdS boundary yields the stress tensor \eqn{divergent}.

%%%%%%%%%%%%%%%%%%%%%%%%%%%%%%%%%%%%%%%%%%%%%%%%
\noindent
{\bf 5. Backreaction.}
%\label{back}
%%%%%%%%%%%%%%%%%%%%%%%%%%%%%%%%%%%%%%%%%%%%%%%%
So far we have used holography to determine the stress tensor of strongly-coupled quantum matter in a fixed metric \eqn{dh2_K}. We will now determine the backreaction of this stress tensor on the metric at leading order in the gravitational coupling $\mathcal{G}$. This means that we will solve the four-dimensional,  semiclassical Einstein equations 
\be
\label{EE}
R_{\mu\nu} - g_{\mu\nu} R = 8\pi G \langle T_{\mu\nu} \rangle
\ee
under the assumption that $g_{\mu\nu}$ is a small,  
\mbox{$O(\mathcal{G})$-perturbation} with respect to the metric \eqn{dh2_K}. Moreover, we will focus on the behaviour near the singularity, so we will assume that the stress tensor takes the form \eqn{divergent}. Thus, in this section the only role of holography is to have determined the right-hand side of \eqn{EE}. The rest is a purely four-dimensional calculation at the boundary.

Let us write $g = g_0 + \lambda h$, with $g_0$ given by \eqn{dh2_K}  and   
%$\lambda \equiv 8\pi \mathcal{G} \Lambda^2$. 
\be
\label{small}
\lambda \equiv 8\pi \mathcal{G} \Lambda^2 = \Lambda^2 / M_{\rm{sp}}^2\,.
\ee
Substituting in \eqn{EE} and expanding to linear order in $\lambda$ leads to a set of coupled, ordinary differential equations in the time variable for the components of $h_{\mu\nu}$. A detailed analysis \cite{prep} shows that, up to $O(\lambda^2)$-corrections and for sufficiently small $|t|$, the  corrected solution 
$g$ can again be written as in \eqn{dh2_K} but with exponents that now obey the modified relations
\begin{equation}
\label{corrected}
 \sum_{i=1}^3 p_i=1 + \lambda 
 %+  O(\lambda^2) 
 \sac \,\,\,\, \sum_{i=1}^3 p_i^2=1 
 %+ O(\lambda^2)
 \,.
\end{equation}

%%%%%%%%%%%%%%%%%%%%%%%%%%%%%%%%%%%%%%%%%%%%%%%%
\noindent
{\bf 6. Discussion.}
%\label{regime}
%%%%%%%%%%%%%%%%%%%%%%%%%%%%%%%%%%%%%%%%%%%%%%%%
We have used holography to compute the stress tensor of strongly-coupled, four-dimensional matter in a fixed Kasner metric \eqn{dh2_K}. All the quantum field theoretical effects near the singularity are included  \footnote{See \cite{Emparan:2021yon} for related work in three dimensions.}. As this is approached, the quantum fields are driven arbitrarily far away from equilibrium and the stress tensor takes the simple form \eqn{divergent}. We emphasize that this does not follow from symmetry considerations alone, in particular from conformal invariance, but is a dynamical property of QFTs with a gravity dual. The fact that it can be derived from linear perturbations around the vacuum geometry \eqn{vacFG} suggests that it is a universal property common to all such theories. 

The near-equilbrium, hydrodynamic description ceases to be valid when the largest expansion rate among \eqn{exp}  becomes comparable to the local temperature. We have verified this even for metrics with parametrically different Kasner exponents \mbox{$p_2 \simeq p_3 \simeq 10^{-3}, p_1\simeq 1-10^{-3}$}. The intuition behind this is that the inverse local temperature controls the equilibration time of the system, so equilibrium cannot be maintained  when the system is forced to expand or contract faster than this time scale. In view of \eqn{exp} and \eqn{TTT} this happens at a time $t\sim \Lambda^{-1}$. Requiring that the spacetime curvature \eqn{curv} at this time be below the species cut-off implies 
$\Lambda < M_{\rm{sp}}$. This means that the matter backreaction, controlled by \eqn{small}, cannot be large, consistently with our perturbative treatment. 

We emphasize that the stress tensor \eqn{divergent} is fundamentally different from that of classical matter in a Kasner spacetime  \cite{Belinski:1973zz,YM_fields,Damour:2002tc}. Classical Maxwell or Yang-Mills fields preserve the BKL chaotic dynamics and do not modify the relations \eqn{exponentssum}, in contrast with \eqn{corrected}. Classical scalar fields lead to an isotropic stress tensor, in contrast with \eqn{divergent}. There is however an interesting similarity between the classical scalar and the quantum result: in both cases the  relations \eqn{exponentssum} are modified in a way that allows all three Kasner exponents to be positive. In the classical case this is well known to replace the BKL chaotic behaviour by a power-law, Kasner-type behaviour near the singularity. This suggests that the same may be true in the presence of quantum matter. To verify this we should extend our analysis to include inhomogeneous perturbations. We expect this to be feasible given the simplicity of the vacuum solution \eqn{vacEF} controlling the dynamics near the singularity.

%%%%%%%%%%%%%%%%%%%%%%%%%%%%%%%%%%%%%%%%%%%%%%%%
\noindent
{\bf Acknowledgements.}
%\label{Acknowledgements}
%%%%%%%%%%%%%%%%%%%%%%%%%%%%%%%%%%%%%%%%%%%%%%%%
We thank Pablo Bueno, Roberto Emparan, Bartomeu Fiol, Sean Hartnoll, Robie Hennigar and Enric Verdaguer for discussions. We are grateful to Roberto Emparan for comments on the manuscript and to Yago Bea for sharing his results on the thermodynamics of the theory. We are supported by grants 2021-SGR-872, CEX2019-000918-M, PID2019-105614GB-C21,  PID2019-105614GB-C22, and PID2022-136224NB-C22 funded by MCIN/AEI/ 10.13039/501100011033/FEDER, UE. This project has received funding from the European Research Council (ERC) under the European Union's Horizon 2020 research and innovation programme (grant number: 101089093 / project acronym: High-TheQ). Views and opinions expressed are however those of the authors only and do not necessarily reflect those of the European Union or the European Research Council. Neither the European Union nor the granting authority can be held responsible for them.

\bibliographystyle{apsrev4-1}
\bibliography{refs_Quantum_Matter_near_a_Cosmological_Singularity}

\end{document}